\documentclass[aps,pre,twocolumn,superscriptaddress,showpacs]{revtex4}
\usepackage{amssymb}
\usepackage{epsfig}
\usepackage{amsmath}
\usepackage{times}

\begin{document}

\title{Nonstationary pattern in unsynchronizable complex networks}
\author{Xingang Wang}
\affiliation{Temasek Laboratories, National University of Singapore, 117508 Singapore}
\affiliation{Beijing-Hong Kong-Singapore Joint Centre for Nonlinear \& Complex Systems
(Singapore), National University of Singapore, Kent Ridge, 119260 Singapore}
\author{Meng Zhan}
\affiliation{Wuhan Institute of Physics and Mathematics, Chinese Academy of Sciences,
Wuhan 430071, China}
\author{Shuguang Guan}
\affiliation{Temasek Laboratories, National University of Singapore, 117508 Singapore}
\affiliation{Beijing-Hong Kong-Singapore Joint Centre for Nonlinear \& Complex Systems
(Singapore), National University of Singapore, Kent Ridge, 119260 Singapore}
\author{Choy Heng Lai}
\affiliation{Department of Physics, National University of Singapore, 117542 Singapore}
\affiliation{Beijing-Hong Kong-Singapore Joint Centre for Nonlinear \& Complex Systems
(Singapore), National University of Singapore, Kent Ridge, 119260 Singapore}

\begin{abstract}
Pattern formation and evolution in unsynchronizable complex
networks are investigated. Due to the asymmetric topology, the
synchronous patterns formed in complex networks are irregular and
nonstationary. For coupling strength immediately out of the
synchronizable region, the typical phenomenon is the on-off
intermittency of the system dynamics. The patterns appeared in
this process are signatured by the coexistence of a giant cluster,
which comprises most of the nodes, and a few number of small
clusters. The pattern evolution is characterized by the giant
cluster irregularly absorbs or emits the small clusters. As the
coupling strength leaves away from the synchronization bifurcation
point, the giant cluster is gradually dissolved into a number of
small clusters, and the system dynamics is characterized by the
integration and separation of the small clusters. Dynamical
mechanisms and statistical properties of the nonstationary pattern
evolution are analyzed and conducted, and some scalings are newly
revealed. Remarkably, it is found that the few active nodes, which
escape from the giant cluster with a high frequency, are
independent of the coupling strength while are sensitive to the
bifurcation types. We hope our findings about nonstationary
pattern could give additional understandings to the dynamics of
complex systems and have implications to some real problems where
systems maintain their normal functions only in the
unsynchronizable state.
\end{abstract}

\date{\today }
\maketitle

\section{Introduction}

Synchronization of complex networks has aroused many interest in
nonlinear science since the discoveries of the small-world and
scale-free properties in many real and man-made systems
\cite{NETWORK:REVIEW,NETSYN:REVIEW}. In this study, one important
issue is to explore the inter-dependent relationship between the
collective behaviors of the complex systems and their underlying
topologies. In particular, many efforts have been paid to the
construction of optimal networks, and a number of factors which
have important affections to the synchronizability of complex
networks have been gradually disclosed. Now it is known that
random networks, due to their small average distances, are
generally more synchronizable than regular networks
\cite{WC:2002,NETSYN:NISHIKAWA}; and scale-free networks, with
weighted and asymmetric couplings, can be more synchronizable than
homogeneous networks \cite{NETSYN:MOTTER,NETSYN:BOCCA}. In these
studies, the standard method employed for synchronization analysis
is the master stability function (MSF), where the network
synchronizability is estimated by an eigenratio calculated from
the coupling matrix, and system which has a smaller eigenratio is
believed to be more synchronizable than that of larger eigenratio
\cite{MSF}. Inspired by this, to improve the network
synchronizability, the only task seems to be upgrading the
coupling matrix so as to decrease the eigenratio, either by
changing the network topology
\cite{NETSYN:NISHIKAWA} or by adjusting the coupling scheme \cite%
{NETSYN:MOTTER,NETSYN:BOCCA}.

The MSF\ method, while bringing great convenience to the analysis,
overlooks the temporal, local property of the system and reflects
only partial information about the system dynamics. Specifically,
from MSF\ we only know ultimately whether the network is globally
synchronizable or unsynchronizable, but do not know how the global
synchronization is reached if the network is synchronizable, or
what's the pattern and how it evolves if the network is
unsynchronizable. These evolution details, or the transient
behavior in system development, contain rich information about the
system dynamics and may give additional insights to the
organization of complex systems. For instance, the recent studies
about synchronization transition have shown that, in the
unsynchronizable states, heterogeneous networks are more
synchronizable (have a higher degree of coherence) than
homogeneous networks at small couplings, while at larger couplings
the opposite happens \cite{SYN:ERSFN}. This crossover phenomenon
of network synchronizability are difficult to understood if we
only look at the final state of the system, but are
straightforward if we look at the transient behaviors of their
evolutions \cite{SYN:ERSFN}. Besides revealing the synchronization
mechanisms, the transient behavior of network synchronization can
also be used to detect the topological scales and hierarchical
structures in the real systems, e.g., the detection of cluster
structures in social and biological networks
\cite{SYN:DETECT1,SYN:DETECT2}. However, despite of its
theoretical and practical significance, the study of transient
dynamics of complex networks is still at its infancy and many
questions remain open, say for example, the pattern evolution of
unsynchronizable complex networks.

Pattern formation in unsynchronizable but near-synchronization
networks has been an important issue in studying the collective
behavior of regular networks \cite{SYN:PARTIAL,DESYN}. By setting
the coupling strength nearby the synchronization bifurcation
point, the system state shares both the dynamical properties of
the synchronizable and unsynchronizable states: a state of high
coherence but is not synchronized. The bifacial dynamical property
makes this state a natural choice in investigating the transition
process of networks synchronization. Previously studies about
regular networks, say for example the lattices \cite{SYN:PARTIAL},
have shown that, when the coupling strength is slightly out of the
synchronizable region, although global synchronization is
unreachable, nodes are still synchronized in a partial sense. That
is, nodes are self-organized into a number of synchronous
clusters. The distribution of these clusters, also called the
synchronous pattern, is determined by a set of factors such as the
coupling strength, the system size and the coupling scheme. As the
coupling strength leaves away from the bifurcation point, the
pattern structure becomes more and more complicated and the system
coherence will be decreased, and finally reaches the turbulence
state. It is worthy of note that the patterns arisen in regular
networks have two common properties: spatially symmetric and
temporally stationary. More specifically, the contents of each
cluster are fixed and the clusters are of translation symmetry in
space. For this reason, we say that \emph{the patterns formed in
regular networks are symmetric and stationary}. These two
properties, as have been discussed in the previous studies
\cite{SYN:PARTIAL,DESYN}, are rooted in the symmetric topology of
the regular networks. This makes it interesting to ask the
following question: how about the patterns in unsynchronizable
complex networks?

Different to the regular networks, in complex networks we are not able to
find any symmetry from their topologies. The asymmetric topology, according
to the pattern analysis developed in studying regular networks \cite%
{SYN:PARTIAL}, will induce two significant changes to the
patterns: 1) the synchronous clusters, if they exist, will be
asymmetric; and 2) all the possible patterns, including the one of
global synchronization, are linearly unstable under small
perturbations. In other words, \emph{the patterns formed in
complex networks are expected to be asymmetric and nonstationary.}
Our mission of this paper is just to understand and characterize
the nonstationary patterns arisen in the development of complex
networks. Specifically, we are trying to investigate the following
questions: 1) is there any pattern arises during the system
evolution? 2) the pattern is stationary or nonstationary? if
nonstationary, how is it evolving and how is it reflected from the
system dynamics? 3) What happens to the pattern properties during
the transition of network synchronization? and 4) How the coupling
strength and bifurcation type affect the pattern properties? By
investigating these dynamical and statistical properties, we wish
to have a global understanding to the dynamics of unsynchronizable
complex networks.

Our main findings are: 1) for coupling strength immediately
outside of the synchronizable region, the system dynamics
undergoes the process of on-off intermittency. That is, most of
the time the system stays on the global synchronization state (the
"off" state) but, once in a while, it develops into a breaking
state (the "on" state) which is composed by a giant cluster and a
few number of small clusters (hereafter we call it the
giant-cluster state). As the system develops, the giant cluster
changes its shape by absorbing or emitting the small clusters,
leading to the "off" or "on" states, respectively; 2) the few
active nodes which escape from the giant cluster with the high
frequencies are coupling-strength independent but are
bifurcation-type dependent. That is, in the neighboring region of
a fixed bifurcation point, the locations of these active nodes do
not change with the coupling strength; if we change the coupling
strength from nearby another bifurcation point (the two
bifurcation points will be explained later), their locations will
be totally changed; 3) as coupling strength leaves away from the
bifurcation point, the giant cluster is gradually dissolved and
more small clusters are generated from it. Eventually, the giant
cluster disappears and the pattern is composed by only the small
clusters (hereafter we call it the scattering--cluster state).
During the course of system evolution, each small cluster may
either increase its size by integrating with other small clusters
or decrease its size by breaking to even small clusters, but it
can never reach to the global synchronization state; 4) besides
the giant cluster, the giant- and scattering-cluster states are
also distinct in their small clusters. For giant-cluster state the
size of the small clusters follows a power-law distribution, while
for scattering-cluster state it follows a Gaussian distribution.

The rest of the paper is going to be arranged as follows. In Sec.
II we will give our model of coupled map network and, based on the
method of MSF, point out the two bifurcation points and the
transition areas that we are going to study with. In Sec. III we
will employ the method of finite-time Lyapunov exponent to predict
and describe the intermittent system dynamics in the bifurcation
regions. Direct simulations about on-off intermittency will be
presented in Sec. IV. By introducing the method of temporal phase
synchronization, in Sec. V we will investigate in detail the
dynamical and statistical properties of the nonstationary pattern.
Meanwhile, properties of the giant- and scattering states will be
compared and the transition between the two states will be
conducted. In Sec. VI we will discuss the phenomenon of active
nodes and investigate their dependence to the network properties.
Discussions and conclusions about pattern evolution in complex
networks will be presented in Sec. VII.

\section{Coupled map networks and the two bifurcation points}

Our model of coupled map network is of the following form%
\begin{equation}
\mathbf{x}_{i}(t+1)=\mathbf{F}(\mathbf{x}_{i}(t))-\varepsilon \sum_{j}G_{i,j}%
\mathbf{H}\left[ \mathbf{f}(\mathbf{x}_{j}(t))\right] .  \label{NET}
\end{equation}%
where $\mathbf{x}_{i}(t+1)=\mathbf{F}(\mathbf{x}_{i}(t))$ is a $d$%
-dimensional map representing the local dynamics on node $i$, $\varepsilon $
is a global coupling parameter, $G$ is Laplacian matrix representing the
couplings, and $\mathbf{H}$ is a coupling function. To facilitate our
analysis, we adopt the following coupling scheme \cite{WLL:2006}:%
\begin{equation}
G_{i,j}=-\frac{A_{i,j}k_{j}^{\beta }}{\sum_{j=1}^{N}A_{i,j}k_{j}^{\beta }},
\label{CM}
\end{equation}%
for $j\neq i$ and $G_{i,i}=1$, with $k_{i}$ the degree of node $i$ and $A\ $%
the adjacent matrix of the network: $A_{i,j}=1$ if node $i$ and
$j$ are connected and $A_{i,j}=0$ otherwise. In comparison with
the traditional coupling schemes, one important advantage we
benefit from this coupling scheme is that the synchronizability of
the network, i.e. the eigenratio of the coupling matrix described
in Eq. [\ref{CM}], can be easily adjusted by the parameter $\beta
$, while the network topology is kept unchanged. This advantage
brings many convenience in network selection since for a given
network topology, even though it is unsynchronizable under the
traditional schemes, can now be synchronizable by adjusting $\beta
$ in Eq. [\ref{CM}]. This convenience is of particular importance
when our studies of network dynamics are focused on the
bifurcation regions, where the network synchronizability should be
deliberately arranged in order to demonstrate
both the two types of bifurcations. We note that the adoption of Eq. [\ref%
{CM}] is only for the purpose of convenient analysis, the findings we are
going to report are general and can also be observed by other coupling
schemes given the network is properly prepared. In practice, we use logistic
map $\mathbf{F}(x)=4x(1-x)$ as the local dynamics and adopt $\mathbf{H}(%
\mathbf{x})=x$ as the coupling function.

We first locate the two bifurcation points of global synchronization. The
linear stability of the global synchronization state $\left\{
x_{i}(t)=s(t),\forall i\right\} $ is determined by the corresponding
variational equations, which can be diagonalized into $N$ blocks of form
\begin{equation}
y(t+1)=\left[ \mathbf{DF}(s)+\sigma \mathbf{DH}(s)\right] y(t),
\label{linearstability}
\end{equation}%
with $\mathbf{DF(s)}$ and $\mathbf{DH}(s)$ the Jacobian matrices of the
corresponding vector functions evaluated at $s(t)$, and $y$ represents the
different modes that are transverse to the synchronous manifold $s(t)$. We
have $\sigma (i)=\varepsilon \lambda _{i}$ for the $i$th block, $i=1,2,...,N$%
, and $\lambda _{1}=0\leq \lambda _{2}\leq ...\leq \lambda _{N}$ are the
eigenvalues of matrix $G$. The largest Lyapunov exponent $\Lambda (\sigma )$
of Eq. [\ref{linearstability}], known as the master stability function (MSF)
\cite{MSF}, determines the linear stability of the synchronous manifold $%
s(t) $. In particular, the synchronous manifold is stable if only $\Lambda
(\varepsilon \lambda _{i})<0$ for each $i=2,...,N$. The set of Lyapunov
exponents $\Lambda (\varepsilon \lambda _{i})$ govern the stability of the
synchronous manifold in the transverse spaces, and a positive value of $%
\Lambda (\varepsilon \lambda _{i})$ represents the loss of the
stability in the transverse space of mode $i$. It was found that
for a large class of chaotic systems, $\Lambda (\sigma )<0$ is
only fulfilled within a limit range in the parameter space $\sigma
\in $ $\left( \sigma _{1},\sigma _{2}\right) $. This indicates
that, to make the global synchronization state linearly stable,
all the eigenvalues $\lambda _{i}$ should be contained within
range $(\sigma _{1},\sigma _{2})$, i.e., $\lambda _{N}/\lambda
_{2}<\sigma _{2}/\sigma _{1}$. For the logistic map employed here,
it is not difficult to prove that $\sigma _{1}=0.5$ and $\sigma
_{2}=1.5$. Therefore, to achieve global synchronization, the
coupling matrix $G$ should be designed with eigenratio $R\equiv
\lambda _{N}/\lambda _{2}<\sigma _{2}/\sigma _{1}=3=R_{c}$.

\begin{figure}[tbp]
\begin{center}
\epsfig{figure=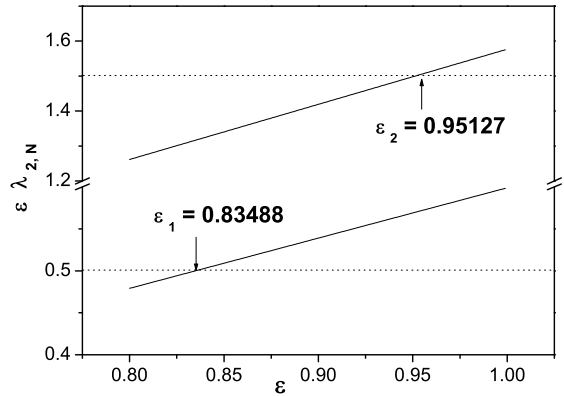,width=0.9\linewidth}
\end{center}
\caption{For scale-free network of $N=1000$ nodes and of average degree $%
<k>=8$, a schematic plot on the generation of the two bifurcation
points as a function of the coupling strength. The long-wave
bifurcation occurs at about $\protect\varepsilon _{1}\approx
0.83488$ which is determined by the condition $\protect\varepsilon
\protect\lambda _{2}=\protect\sigma _{1}=0.5$
(the lower line). The short-wave bifurcation occurs at about $\protect%
\varepsilon _{2}\approx 0.95127$ which is determined by the condition $%
\protect\varepsilon \protect\lambda _{N}=\protect\sigma _{2}=1.5$ (the upper
line).}
\end{figure}

Besides the condition of $R<R_{c}$, to guarantee the synchronization, we
also need to set the coupling strength in a proper way: either small or
large couplings may deteriorate the synchronization. If $\varepsilon
<\varepsilon _{1}=\sigma _{1}/\lambda _{2}$, the couplings are too weak to
restrict the node trajectories to the synchronous manifold; while if $%
\varepsilon >\varepsilon _{2}=\sigma _{2}/\lambda _{N}$, the couplings will
be too strong and actually act as large perturbations to the synchronization
manifold. Therefore, to achieve the global synchronization, we also require $%
\varepsilon _{1}<\varepsilon \,<\varepsilon _{2}$. The two critical
couplings $\varepsilon _{1}$ and $\varepsilon _{2}$, which are named as the
long-wave (LW) \cite{BIF:SHORT} and short-wave (SW) bifurcations \cite%
{BIF:FAST} respectively in the studies of regular networks, thus stand as
the boundaries of the synchronizable region. Our studies about network
synchronization will be focused on the neighboring regions of the two
bifurcation points, i.e., the region of $\varepsilon \lesssim \varepsilon
_{1}$ or $\varepsilon \gtrsim \varepsilon _{2}$.

By the standard BA growth model \cite{NETWORK:REVIEW}, we construct a
scale-free network of $10^{3}$ nodes and of average degree $\left\langle
k\right\rangle =8$. By setting $\beta =2.5$ in Eq. [\ref{CM}], we have $%
\lambda _{2}\approx 0.6$ and $\lambda _{N}\approx 1.58$. Because of $%
R=\lambda _{N}/\lambda _{2}\approx 2.6<R_{c}$, the network is globally
synchronizable. Also, because of $\lambda _{2}>\sigma _{1}$ and $\lambda
_{N}>\sigma _{2}$, both the two bifurcations can be realized by adjusting
the coupling strength within range $\varepsilon \in (0,1)$. In specific,
when $\varepsilon <\varepsilon _{1}\approx 0.835$, we have $\varepsilon
\lambda _{2}<\sigma _{1}$ and $\varepsilon \lambda _{N}<\sigma _{2}$, the
synchronous manifold loses its stability at the lower boundary of the
synchronizable region and LW bifurcation occurs; and when $\varepsilon
>\varepsilon _{2}\approx 0.95$, we have $\varepsilon \lambda _{2}>\sigma
_{1} $ and $\varepsilon \lambda _{N}>\sigma _{2}$, the synchronous manifold
loses its stability at the upper boundary of the synchronizable region and
SW bifurcation occurs [Fig. 1]. In the following we will fix the network
topology and the parameter $\beta $, while generating the various patterns
by changing the coupling strength $\varepsilon $ nearby the two bifurcation
points.

\section{Finite-time Lyapunov exponent}

Before direct simulations, we first give a qualitative description
(prediction) on the possible system dynamics in bifurcation
regions. To concrete our analysis, in the following we will only
discuss the situation of SW bifurcation ($\varepsilon \lesssim
\varepsilon _{1}$), while noting
that the same phenomena can be found at the LW bifurcation as well ($%
\varepsilon \gtrsim \varepsilon _{2}$) . In preparing the
unsynchronizable states, we only let $\Lambda (\lambda _{2})$ be
slightly puncturing into the unstable region, while keeping all
the other exponents still staying in the stable region, i.e.,
$\Lambda (\lambda _{2})\gtrsim 0$ and $\Lambda (\lambda _{i})<0$
for $i=3,...N$. With this setting, the synchronous manifold is
only desynchronized in the transverse space of mode $2$. As such,
the system possesses only two positive Lyapunov exponents, one is
$\Lambda (\lambda _{0})$ which is associated to the synchronous
manifold itself and another one is $\Lambda (\lambda _{2})$.
Noticing that $\Lambda (\lambda )$ are asymptotic averages, and,
as so, they account only for the global stability properties, but
do not warrant the possible coherent sets arising in the system
evolutions. These coherent sets, for regular networks, refer to
the stationary, symmetric patterns to which the system finally
develops. While for complex networks, these sets can be the
temporal, irregular clusters emerged in the process of system
evolution.

In the region of $\varepsilon \lesssim \varepsilon _{1}$, although
global synchronization is unreachable, the system may still keep
with the high coherence due to the existence of the synchronous
clusters. Especially, there could be some moments at which all the
trajectories are restrained to a small region in the phase space,
very close to the situation of global synchronization. This
varying system coherence, however, can not be reflected from the
asymptotic value $\Lambda (\lambda )$. To characterize the
variation, we need to employ some new quantities which are able to
capture the temporal behavior of system. One of such quantities is
the finite-time Lyapunov exponent (FLE), a technique developed in
studying chaos transition in nonlinear science \cite{FLE}. In
stead of asymptotic average, FLE measures the diverging rate of
nearby trajectories only in a finite time interval $T$.
\begin{equation}
\Lambda _{i}=\frac{1}{T}\sum_{t=(i-1)T}^{iT}\ln \mathbf{DH}(s(t)).
\end{equation}%
As our studies are focused on the situation of one-mode desynchronization,
the stability of the synchronous manifold and the temporal behavior that it
displays are therefore expected to be more reflected from the variation of $%
\Lambda _{2,i}$, the FLE that associates with mode $2$. With $\varepsilon
=0.83$ and $T=100$, we plot in Fig. 2 the time evolution of $\Lambda _{2,i}$%
. It is found that, although with a positive asymptotic value about $%
\left\langle \Lambda _{2,i}\right\rangle \approx 6\times 10^{-3}$, the
instant value of $\Lambda _{2,i}$ penetrates to the negative region at a
high frequency. According to the different signs of $\Lambda _{2,i}$, the
system evolution is divided into two types of intervals: the divergent
interval and the contractive interval. In the divergent intervals we have $%
\Lambda _{2,i}>0$ and the system dynamics is temporarily dominated
by the divergence of the node trajectories from the synchronous
manifold; while in the contractive intervals we have $\Lambda
_{2,i}<0$ and the system dynamics is temporarily dominated by the
convergence of the node trajectories to the synchronous manifold.

\begin{figure}[tbp]
\begin{center}
\epsfig{figure=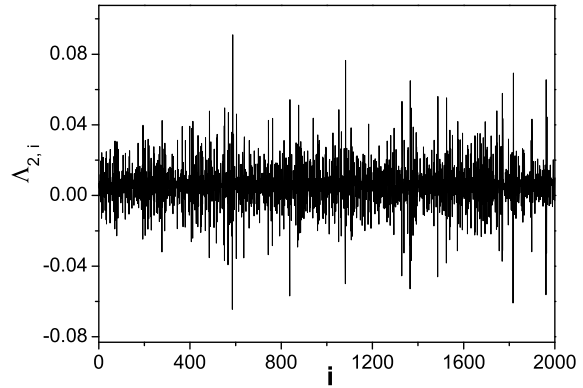,width=0.9\linewidth}
\end{center}
\caption{For $\protect\varepsilon =0.83$ in Fig. 1, the time
evolution of the finite-time Lyapunov exponent $\Lambda _{2,i}$
calculated on intervals of length $T=100$. It is observed that,
while having the positive asymptotic
value $\left\langle \Lambda _{2,i}\right\rangle >0$, the temporal value of $%
\Lambda _{2,i}$ is penetrating into the negative region frequently.}
\label{fig:fle}
\end{figure}

The variation of $\Lambda _{2,i}$, reflected on the process of
pattern evolution, characterizes the travelling property of the
system dynamics among the neighboring regions of two different
kinds of states: the desynchronization state and the
synchronization state. In Fig. 2, the minimum value of $\Lambda
_{2,i}$ is about $-0.07$, during this contractive interval the
node trajectories will converge to the synchronous manifold by an
amount of $e^{\min \Lambda _{2,i}T}\approx e^{-7}\approx 10^{-3}$
on average. Assuming that before entering this interval the
average distance
between the node trajectories is $\Delta $ (for logistic map we always have $%
\Delta <1$), then at the end of this interval the average distance
is decreased to $\Delta \times 10^{-3}$, a small value which is
usually overshadowed by\ noise in practice. Due to this small
distance, the system can be practically regarded as already
reached the synchronization state. On the other hand, if the
system enters a divergent interval, the node trajectories will
diverge from each other and, at the end of this interval, their
average distance will be increased by an order of $10^{3}$. This
large distance will deteriorate the ordered trajectories (or the
high coherence of the system dynamics) that achieved during the
contractive intervals, and leading to the incoherent, breaking
state. The pattern of the breaking state, however, is not unique.
Depending on the initial conditions and the divergence intervals,
the pattern may assume the different configurations. Therefore,
based on the observations of $\Lambda _{2,i}$ [Fig. 2] the
dynamics of unsynchronizable networks can be intuitively
understood as an intermittent travelling among the synchronization
state and the different kinds of desynchronization states.

\section{On-off intermittency described by complete synchronization}

We now investigate the system dynamics by direct simulations. To implement,
we first prepare the system to be staying on the synchronization state. This
can be achieved by adopting a large coupling strength from the
synchronizable region, i.e. $\varepsilon _{1}<\varepsilon <\varepsilon _{2}$%
. After synchronization is achieved, we then decrease $\varepsilon $ to a
value slightly below the bifurcation point $\varepsilon _{1}$ and, in the
meantime, an instant small perturbation is added on each node. In practice,
we take i.i.d (independent identically distributed) noise of strength $%
1\times 10^{-5}$ as the perturbations. After this, we release the
system and let it develop according to Eq. (\ref{NET}). Since
$\varepsilon <\varepsilon _{1}$, the synchronization state is
unstable and, triggered by the noises, the node trajectories begin
to diverge from each other. The divergent trajectories, however,
will frequently visit the neighborhood of the synchronous
manifold, especially during those contractive intervals of small
$\Lambda _{2,i}$ [Fig. 2]. The intermittent system dynamics is
plotted in
Fig. 3(a), where the average trajectory distance $\Delta X=\frac{1}{N}%
\sum_{i=1}^{N}x_{i}-\vec{x}$ is plotted as a function of time. As we have
predicted from LLE, the system dynamics indeed undergoes an intermittent
process. To characterize the intermittency, we plot in Fig. 3(b) the
laminar-phase distribution of the $\Delta X$ sequence plotted in Fig. 3(a).
It is found that the laminar length $\tau $ (the time interval between two
adjacent bursts of amplitude $\Delta X(t)>10^{-3}$) and the probability $%
p(\tau )$ for it to appear follow a power-law scaling $p(\tau )\sim \tau
^{-\gamma }$. The fitted exponent is about $\gamma \approx -1.5\pm 0.05$,
with a fat tail at large $\tau $ due to the finite simulating time.

\begin{figure}[tbp]
\begin{center}
\epsfig{figure=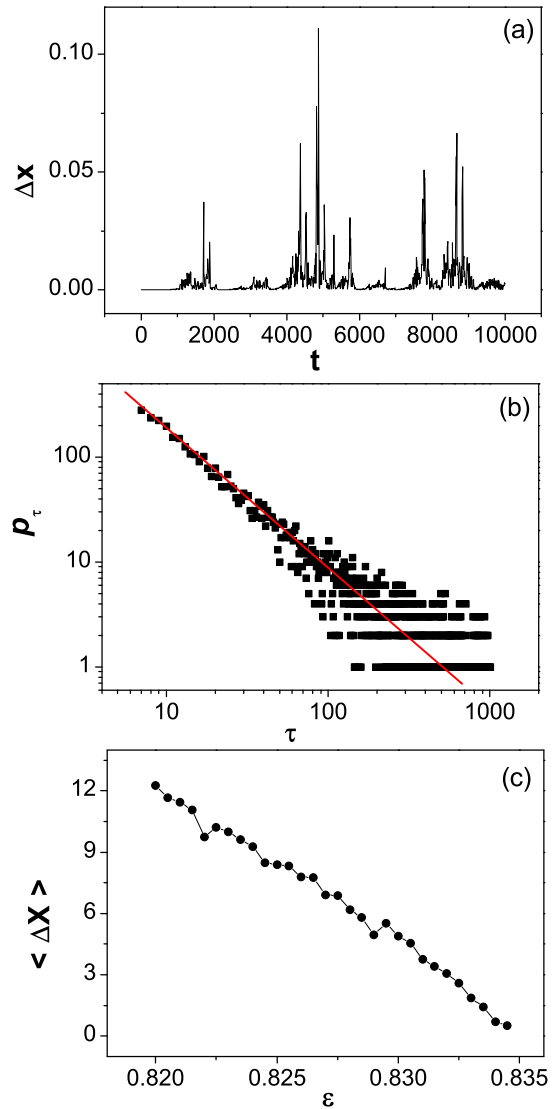,width=0.9\linewidth}
\end{center}
\caption{(Color online) The on-off intermittency of the system dynamics
nearby the LW bifurcation at $\protect\varepsilon =0.83$. (a) The time
evolution of the average trajectory distance $\Delta X$. (b) The
laminar-phase distribution of $\Delta X$, which follows a power-law scaling
with the fitted exponent around $3/2$. (c) The transition behavior of the
average distance $\left\langle \Delta X\right\rangle $ nearby the LW
bifurcation point $\protect\varepsilon _{1}$, where a linear relation is
found between the two quantities.}
\end{figure}

In chaos theory, intermittent process of laminar-phase exponent
$-3/2$ is classified as the "on-off" intermittency, a typical
phenomenon observed in dynamical systems with a symmetric
invariant set \cite{INTERMITTENT}. On-off intermittency is also
reported in chaos synchronization of regular networks, where the
invariant set refers to the synchronous manifold, and the "off"
state refers to the long stretches that the system dynamics is
staying nearby the synchronous manifold and the "on" state refers
to the short bursts that the system dynamics is staying away from
the synchronous manifold. Therefore, in terms of laminar-phase
distribution, the intermittency we have found in complex networks
[Fig. 3] has no difference to the that of the regular networks,
despite of the drastic difference between their topologies. We
have also investigated the transition behavior of the averaged
distance $\left\langle \Delta X(t)\right\rangle $ nearby the
bifurcation points. As shown in Fig. 3(c), a linear relation between $%
\left\langle \Delta X(t)\right\rangle $ and $\varepsilon $ is found in the
region of $\varepsilon \lesssim \varepsilon _{1}$. This linear transition of
the system performance, again, is consistent with the transition of regular
networks \cite{PRK:2001}. Therefore, in terms of complete synchronization,
the on-off intermittency we have found in complex networks has no difference
to that of the regular networks.

\section{Pattern evolution in complex networks}

To reveal the unique properties of the system dynamics that
induced by the complex topology, we go on to investigate the
pattern formation of unsynchronizable networks by the method of
temporal phase synchronization (TPS).

\subsection{Temporal phase synchronization}

TPS is defined as follows. Let $x_{i}(t)$ be the time sequence
recorded on node $i$, we first transform it into a symbolic
sequence $\theta _{i}(t)$ according to the following equations

\begin{equation}
\theta _{i}(t)=%
\begin{cases}
0, & \text{if }x_{i}(t)<0.5, \\
1, & \text{if }x_{i}(t)\geq 0.5.%
\end{cases}
\label{TRANS}
\end{equation}%
Then we divide $\theta _{i}(t)$ into short segments of the equal
length $n$. Regarding each segment as an new element, we therefore
have transformed the long, variable sequence $x_{i}(t)$ into a
short, symbolic sequence $\Theta _{i}(t^{\prime })$. If at moment
$t^{\prime }$ we have $\Theta _{i}(t^{\prime })=\Theta
_{j}(t^{\prime })$, then we say that TPS is achieved between the
nodes $i$ and $j$. The collection of nodes which have the same
value of $\Theta \ $at moment $t^{\prime }$ are defined as a
temporarily synchronous cluster, and all the synchronous clusters
constitute the temporarily pattern of the system. During the
course of system evolution, the clusters will change their shapes
and contents and the pattern will change its configuration.

In comparison with the method of complete synchronization, the
advantage we benefit from TPS is obvious: it makes the synchronous
pattern detectable. With complete synchronization, it is almost
impossible for two nodes to have exactly the same variable at the
same time. Despite the fact that at some moments the system has
already reached the high-coherence states (formed during those
contractive intervals in Fig. 3(a)), with complete synchronization
we are not able to distinguish these states from those
low-coherence ones quantitatively (formed during those divergent
intervals in Fig. 3(a)). (A remedy to this difficulty seems to
define the clusters by the method of threshold truncation, i.e.,
nodes are regarded as synchronized if the distance between their
trajectories is smaller than some small value. However, this
definition of synchronization will induce the problem of cluster
idenfication, as the same state may generate different patterns if
we choose the different reference nodes.) On the contrary, TPS
focuses on the loose match (phase synchronization) between the
node variables over a period of time. By requiring an exact match
of the discrete variable $\Theta $, the synchronous pattern is
uniquely defined; while by requiring the match of the long
sequences of $\theta $, the "synchronous" nodes are guaranteed
with a strong coherence.

\subsection{Pattern evolution of the giant-cluster state}

With the same set of parameters as in Fig. 3(a), by the method of
TPS we plot in Fig. 4 the time evolutions of two basic quantities
of pattern evolution: the number of synchronous clusters $n_{c}$
and the size of the largest cluster $L_{\max }$. It is found that,
similar to the phenomenon in complete synchronization [Fig. 3(a)],
on-off intermittency is also found in the TPS$\ $quantities
$n_{c}$ and $L_{\max }$. In Fig. 4(a) it is shown that
most of the time the system is broken into only a few number of clusters, $%
n_{c}=2$ or $3$, while occasionally it is broken into a quite large number
of clusters, $10<n_{c}<50$, or united to the synchronization state, $n_{c}=1$%
. The intermittent pattern evolution is also reflected on the sequence of $%
L_{\max }$ [Fig. 4(b)], where most of the time the size of the giant cluster
is about $L_{\max }\approx N$, while occasionally it decreases to some small
values of $L_{\max }<N/2$. As we have discussed previously, the main
advantage we benefit from TPS is in identifying the clusters. This advantage
is clearly shown in Figs. 4(a) and (b), where for any time instant the two
quantities $n_{c}$ and $L_{i}$ are uniquely defined. Besides cluster
identification, we also benefit from TPS in quantifying the synchronization
degrees. In specific, the different coherence states shown in Fig. 3(a) now
can be clearly quantified: high coherence states are those of smaller $n_{c}$
or larger $L_{\max }$. Specially, the synchronization state now is
unambiguously defined as the moments of $n_{c}=1$ in Fig. 4(a)\ or, equally,
the moments of $L_{\max }=N$ in Fig. 4(b).

\begin{figure}[tbp]
\begin{center}
\epsfig{figure=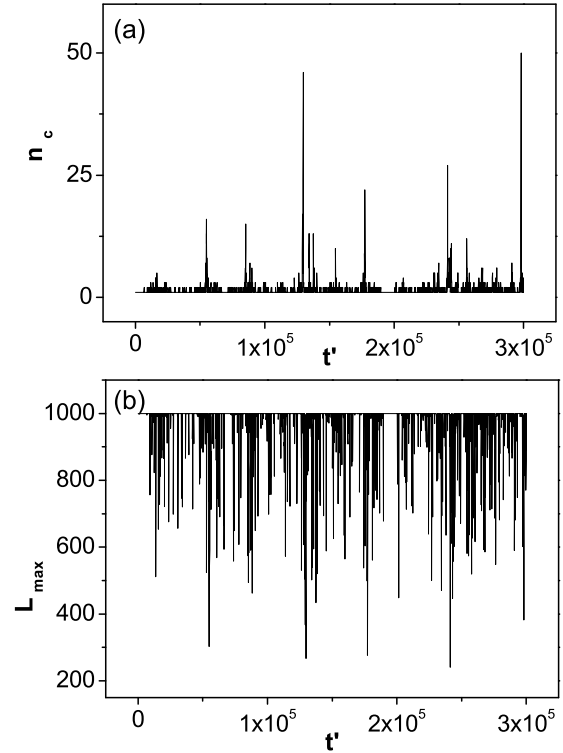,width=0.9\linewidth}
\end{center}
\caption{For the same set of parameters as in Fig. 3(a). The time
evolutions of the TPS quantities. (a) The number of the
synchronous clusters $n_{c}$ and (b) the size of the giant cluster
$L_{\max }$. The synchronization state is defined as the moments
$n_{c}=1$ in (a) or $L_{\max }=N$ in (b).}
\end{figure}

We go on to investigate the pattern evolution by statistical
analysis. The first statistic we are interested is the
laminar-phase distribution of the synchronization state, i.e. the
time intervals that $n_{c}=1$ in Fig. 4(a) or $L_{\max }=N$ in
Fig. 4(b). In its original definition, laminar phase refers to the
time intervals $\tau $ that all node trajectories stays within a
small distance from the synchronous manifold, therefore the actual
value of $\tau $ is varying with the predefined threshold
distance. This uncertainty is overcome in TPS. As shown in Fig.
4(a), in TPS the "off" state refers to the moments of $n_{c}=1$
specifically. The laminar-phase distribution of $n_{c}$ is plotted
in Fig. 5(a). In consistency with the distribution of complete
synchronization [Fig. 3], the laminar-phase distribution of
$n_{c}$ also follows a power-law scaling and has the same exponent
$\gamma \approx -1.5\pm 0.1$. Therefore the use of TPS, while
bringing convenience to the pattern analysis, still capture the
basic properties of the on-off intermittency. The second statistic
we are interested is the size distribution of the largest cluster,
an important indicator for the coherence degree of the system. For
the $L_{\max }$ sequence plotted in Fig. 4(b), in Fig. 5(b) we
plot the size distribution of
$L_{\max }$. It is seen that the probability of finding large cluster $%
L_{\max }\approx N$ is much higher than that of small cluster of $L_{\max
}<500$. In particular, the probability for finding clusters of $L_{\max
}>990 $ is about $20$ percent and for $L_{\max }>990$ it is about $70$
percent. Therefore, in the region of $\varepsilon \lesssim \varepsilon _{1}$%
, the distinct feature of the system patterns is the existence of a giant
cluster. Due to this special feature, we call these states the giant-cluster
state.

Besides the giant cluster, we are also interested in the properties of the
small clusters. We plot in Fig. 5(c) the distribution of $n_{c}$ and in Fig.
5(d) the size distribution of the small clusters $L_{i}$ that surround the
giant cluster in the pattern. As shown in Fig. 5(c), the distribution of $%
n_{c}$ follows a power-law scaling with the fixed exponent is about $\gamma
\approx -3\pm 0.05$. The heterogeneous distribution of $n_{c}$ indicates
that, in the giant-cluster state, the system is usually broken into only a
few number of clusters. An interesting finding exists in the size
distribution of the small clusters. As shown in Fig. 5(d), in range $%
L_{i}\in \lbrack 1,N/2]$ a power-law scaling is found between $P_{Li}$ and $%
L_{i}$, with the fitted exponent is about $\gamma \approx -1.1\pm 0.05$. The
distribution of $L_{i}$ confirms the finding of Fig. 5(c)\ that the small
clusters which join or separate from the giant cluster are usually of small
size.

\begin{figure}[tbp]
\begin{center}
\epsfig{figure=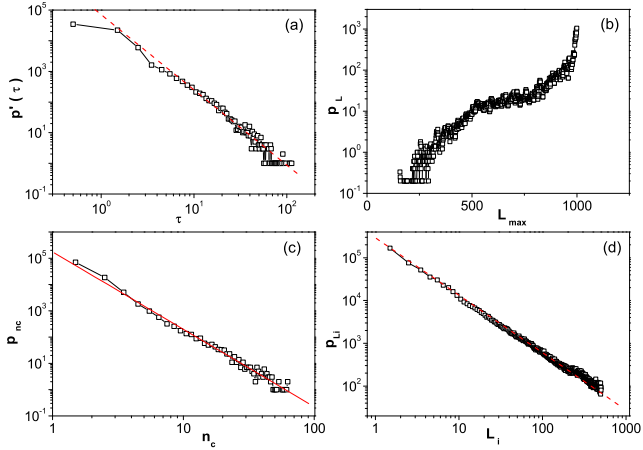,width=\linewidth}
\end{center}
\caption{(Color online) Statistical properties of the on-off intermittency
plotted in Fig. 4. (a) The power-law scaling of the laminar-phase
distribution of $n_{c}$. The fitted slope is about $-2.3\pm 0.05$. (b) The
size distribution of the size of the giant cluster. (c) The power-law
distribution of the number of small clusters $n_{c}$. The fitted slope is
about $-3\pm 0.1$. (d) The power-law scaling on the size distribution of the
small clusters. The fitted slope is about $-1.2\pm 0.01$. }
\end{figure}

Combining the findings of Fig. 4 and Fig. 5, the picture of pattern
evolution in the bifurcation region $\varepsilon \lesssim \varepsilon _{1}$
now becomes clear. Generally speaking, the evolution can be divided into two
opposite dynamical processes happening around the giant cluster: the
separation and integration of the small clusters. During the separation
process, the small clusters are escaped from the giant cluster, which
weakens the dominant role of the giant cluster and makes the pattern
complicated. However, the separated clusters occupy only a small proportion
of the nodes [Fig. 5(c)], the majority nodes are still attached to the giant
cluster, which sustains the synchronization skeleton and keeps the system on
the high coherence states. At some rare moments the giant cluster may
disappears, and the pattern is composed by only small clusters of $%
L_{i}<N/2 $. At these moments, the synchronization skeleton is
broken, the pattern becomes even complicated and the system
coherence reaches its minimum. In contrast, during the process of
cluster integration,\ the giant cluster will increase it size by
attracting the small clusters , and gradually towards the state of
global synchronization. It should be noticed that the separation
and integration processes are uneven and are typically occurring
at the same time. For instance, during the separation process,
while the system evolution is dominated by the separation of new
small clusters from the giant cluster, there could be some small
clusters rejoin to the giant cluster.

\subsection{Pattern evolution of the scattering-cluster state}

As we further decrease the coupling strength from $\varepsilon
_{1}$, the picture of pattern evolution will be totally changed.
With $\varepsilon =0.79 $, we plot in Fig. 6 the same statistics
as in Fig. 5. The first observation is the loss of the global
synchronization state, as can be found from the time variation of
$n_{c}$ plotted in Fig. 6(a). The loss of global synchronization
becomes even clear if we compare Fig. 6(a)\ with Fig. 4(a): in
Fig. 6(a), except the moment at $t=0$, the system can never reach
the
synchronization state at $n_{c}=1$ and very often it is broken into a \emph{%
large} number of small clusters at about $n_{c}\sim $ $10^{2}$. The fact
that the pattern is decomposed into a large number of small clusters is also
manifested by the distribution of $n_{c}$, as plotted in Fig. 6(b). Instead
of the power-law distribution found in the giant-cluster state, in the
scattering-cluster state $n_{c}$ follows a Gaussian distribution [Fig.
6(b)]. As $\varepsilon $ further decreases from $\varepsilon _{1}$, the mean
value of $n_{c}$ will shift to the larger values, as indicated by the $%
\varepsilon =0.78$ curve plotted in Fig. 6(b).\ The second observation is
the disappearance of the giant cluster. As shown in Fig. 6(c), the size
distribution of the largest cluster also follows a Gaussian distribution,
with its mean value locates at $\left\langle L_{\max }\right\rangle <N/2$.
The distribution of Fig. 6(c) is very different to that of Fig. 5(b), where
in Fig. 5(b) the largest (giant) cluster has size $L_{\max }\approx N$ in
most of the time. As $\varepsilon $ decreases, the mean value of the largest
cluster $\left\langle L_{\max }\right\rangle $ will shift to small values
and the variance of $L_{\max }$ will be decreased, as indicated by the $%
\varepsilon =0.78$ curve plotted in Fig. 6(c). Similar to plot of Fig. 5(d),
we have also investigated the distribution of $L_{i}$, the sizes for all the
small clusters appeared in the system evolution [Fig. 6(d)]. It is found
that the distribution of $L_{i}$ follows a power-law distribution for $%
L_{i}<N/2$, while having an exponential tail for $L_{i}>N/2$. Numerically we
find that the exponent of the power-law section, i.e. in range $L_{i}\in
\lbrack 1,200)$, is about $-2\pm 0.05$, while the fitted exponent for the
exponential section is about $-4.5\times 10^{-3}\pm 2\times 10^{-5}$. These
two exponents, however, are changing with $\varepsilon $. As $\varepsilon $
decreases, the two exponents will shift to some small values.

\begin{figure}[tbp]
\begin{center}
\epsfig{figure=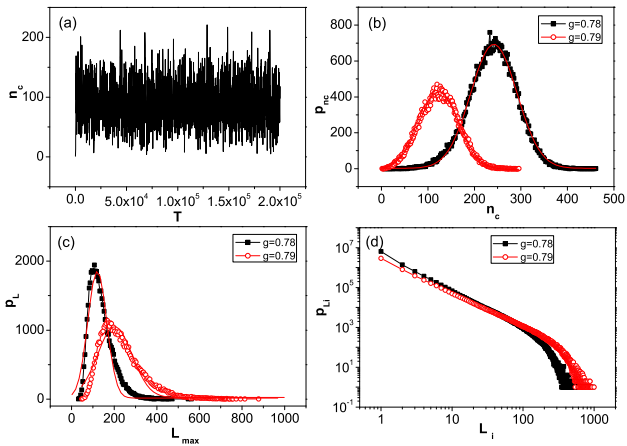,width=\linewidth}
\end{center}
\caption{(Color online) The dynamical and statistical properties of pattern
evolution for $\protect\varepsilon =0.79$. (a) The time evolution of $n_{c}$%
. (b) The Gaussian distribution of the number of the small clusters $n_{c}$.
(b) The Gaussian distribution of the size of the largest cluster $L_{\max }$%
. (d) The two-segment distribution on the size of the small clusters $L_{i}$%
. In the region of $L_{i}<200$, $L_{i}$ follows a power-law distribution
with fitted exponent is about $-2\pm 0.05$; while for $L_{i}>200$, the
distribution is exponential with the fitted exponent is about $-4.5\times
10^{-3}\pm 2\times 10^{-5}$. As $\protect\varepsilon $ further decreases
from $\protect\varepsilon _{1}$, the largest cluster becomes even smaller
and more small clusters are emitted out from it. As illustrated by the $%
\protect\varepsilon =0.78$ curves plotted in (b), (c) and (d).}
\end{figure}

Combining Fig. 5 and Fig. 6, we are able to outline the transition process
of network synchronization nearby the bifurcation points, i.e., the
transition from the giant-cluster state to the scattering-cluster state as $%
\varepsilon $ leaves away from $\varepsilon _{1}$. In the region of $%
\varepsilon \lesssim \varepsilon _{1}$, the pattern is composed by
a giant cluster and a few number of small clusters, i.e. the
giant-cluster state. As $\varepsilon $ decreases from $\varepsilon
_{1}$ gradually, more and more small clusters will be emitted out
from the giant cluster and, as a consequence, both the size of the
giant cluster and the fraction of synchronization time will be
decreased. Then, at about $\varepsilon _{c}\approx 0.832$, the
giant cluster disappears and the pattern of the system is composed
by several larger clusters, of size $L_{\max }\lesssim N/2 $,
together with many small clusters of heterogenous size
distribution, i.e. the scattering-cluster state. After that, as
$\varepsilon $ decreases from$\ \varepsilon _{c}$, the clusters
shrink their size by breaking into even small clusters, and the
pattern becomes even complicated. The detail transition from the
giant-cluster state to the scattering-cluster state is presented
in Fig. 7, where the average number of clusters that the system is
broken into $\left\langle n_{c}\right\rangle $, Fig. 7(a), and the
average size of the largest cluster $\left\langle L_{\max
}\right\rangle $, Fig. 7(b), are plotted as a function of the
coupling strength in the LW bifurcation region. The transition is
found to be smooth and steady, just as we have expected. Besides
the giant cluster, another difference between the giant-cluster
and scattering-cluster states exists in their pattern evolutions.
In the giant-cluster state, while the configuration of the giant
cluster is continuously updated by emitting or absorbing the small
clusters, its main contents are stable and do not change with
time. In contrast, in the scattering-cluster state the small
clusters integrate with or separate from each other in a random
fashion. Although occasionally there could be some large clusters
show up in the pattern of the scattering-cluster state [Fig.
6(d)], these "large" clusters, however, are very fragile and will
break into small clusters again in a short time. This
quick-dissolving property stops the scattering-cluster state from
having a high coherence.

\begin{figure}[tbp]
\begin{center}
\epsfig{figure=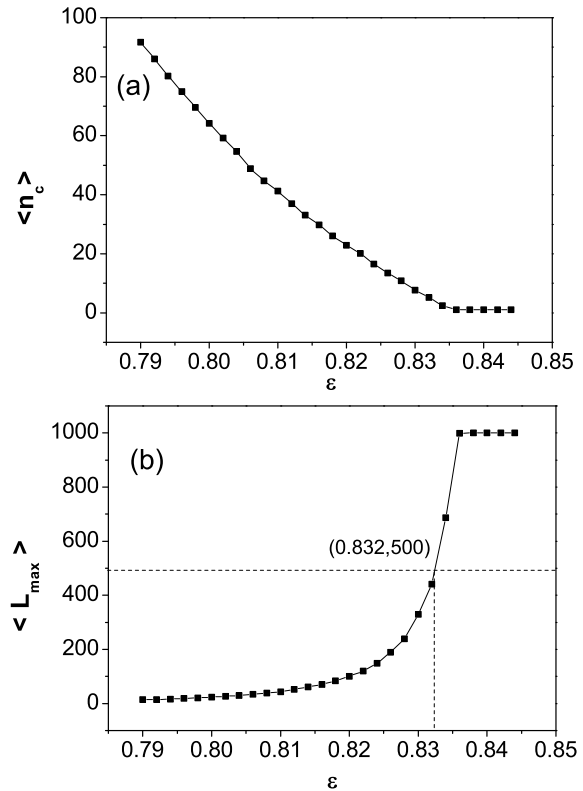,width=0.9\linewidth}
\end{center}
\caption{The transition process of the network synchronization
nearby the LW bifurcation point $\protect\varepsilon _{1}$. (a)
The average number of clusters that the system is broken into as a
function of coupling strength. (b) The average size of the largest
cluster as a function of coupling strength. Each date is an
averaged result over $10^{8}$ time steps. }
\end{figure}

\section{Characterizing the active nodes}

In the giant-cluster state, most of the nodes are organized into
the giant cluster while few nodes, either in forms of small group
or isolated node, are separating from or joining to the giant
cluster with a high frequency. These active nodes, although are
few in amount, play an important role in network synchronization.
Clearly, a proper characterization of these nodes will deepen our
understandings on the system dynamics and give indications to the
improvement of network performance. For instance, to improve the
synchronizability of the system, we may either remove the few most
active nodes from the network, or update their coupling strengths
specifically.

In characterizing the active nodes, the following properties are
of general interest: 1) what's the dependence of the node activity
on the network topology?\ can we characterize these nodes by the
known network properties such as node degree or betweenness? 2)
are their locations sensitive to the coupling strength? and 3)
what's the effect of bifurcation type on their locations? In the
following we will explore these questions by numerical
simulations.
\begin{figure}[tbp]
\begin{center}
\epsfig{figure=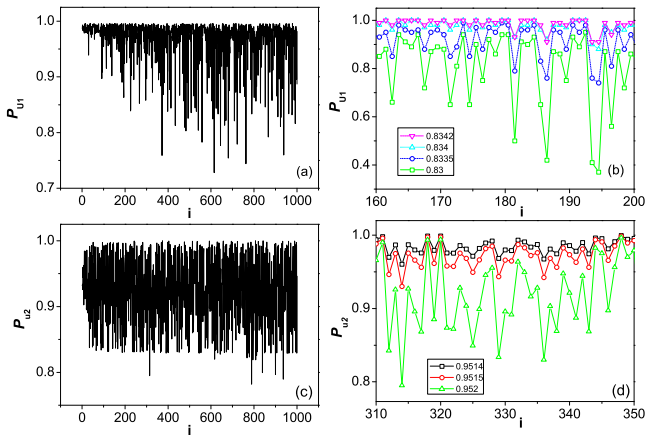,width=\linewidth}
\end{center}
\caption{(Color online) The properties of the active nodes. (a) For the
giant-cluster state shown in Fig. 4, the probability that node stays in the
giant cluster versus the node index. (b) A segment of (a) but with different
coupling strengths nearby the LW bifurcation point $\protect\varepsilon _{1}$%
. (c) For the giant-cluster state ($\protect\varepsilon =0.952$)
nearby the SW bifurcation point, the probability that node stays
in the giant cluster versus the node index. (d) A segment of (c)
under different coupling strengths nearby the SW bifurcation point
$\protect\varepsilon _{2}$.}
\end{figure}

We first try to characterize the active nodes by their topological
properties. For the giant-cluster state described in Fig. 4, we plot in Fig.
8(a)\ the probability $p_{u1}$ that each node stays in the giant cluster.
While the majority nodes stay in the giant cluster with a high probability $%
p_{u1}\approx 1$, few nodes are of unusually small probabilities:
$1$ percent of the nodes have $p_{u1}<0.8$. One important
observation of Fig. 8(a) is that the locations of the active nodes
are entangled with those of the stable nodes. Noticing that in the
BA growth model node of higher index in general assume the smaller
degree, the observation of Fig. 8(a)\ therefore indicates the
independence of the node degree to the node stability, or the
inaccuracy of using degree to characterize the node activity.
Specifically, in Fig. 8(a) the $5$ most unstable nodes, by a
descending order of $p_{u1}$, are those of degrees $k=47$, $36$,
$26$, $10$, $4$, respectively. Except the one of $k=4$, all the
other nodes have higher degrees. Another well-known topological
property of complex network is the node betweenness, which counts
the number of shortest pathes that pass through each node and
actually evaluates the node importance from the global-network
point of view. This global-network property, however, is also
incapable to characterize the active nodes. In Tab. 1 we list the
detail information about the $5$ most active nodes in Fig. 8(a),
where the inaccuracy of node degree or node betweenness in
characterizing the active nodes are summarized.

\begin{center}
\begin{table}[tbh]
\caption{For the attaching probability $p_{i}$ plotted in Fig. 8(a), we list
the $5$ most unstable nodes and try to characterize them by a set of
topological quantities including the node index $i$, the attaching
probability $p_{i}$, the stability rank $p_{i}$\_rank, the node degree $k_{i}
$, the degree rank $k_{i}$\_rank, the node betweenness $B_{i}$, and the
betweenness rank $B_{i}$\_rank.}%
\begin{tabular}{|c|c|c|c|c|c|c|}
\hline
Node index $i$ & $p_{i}$ & $p_{i}$\_rank & $k_{i}$ & $k_{i}$\_rank & $B_{i}$
& $B_{i}$\_rank \\ \hline
615 & 0.72797 & 1 & 5 & 39$\rightarrow $537 & 1301 & 280 \\ \hline
762 & 0.74424 & 2 & 5 & 39$\rightarrow $537 & 1375 & 356 \\ \hline
680 & 0.75416 & 3 & 4 & 1$\rightarrow $338 & 1254 & 680 \\ \hline
372 & 0.7591 & 4 & 6 & 538$\rightarrow $645 & 1440 & 406 \\ \hline
938 & 0.75972 & 5 & 4 & 1$\rightarrow $338 & 1215 & 159 \\ \hline
\end{tabular}%
\label{table_LW}
\end{table}
\end{center}

We go on to investigate the affection of the coupling strength on the
locations of the active nodes. In Fig. 8(b) we fix the network topology and
compare the node activities under different coupling strengths nearby the
bifurcation point $\varepsilon _{1}$. It is found that, despite of the
changes in  $p_{u1}$, the locations of the active nodes are kept unchanged.
That is, the active nodes are always the first ones to escape from the giant
cluster whenever the network is unsynchronizable. We have also investigated
the affection of the bifurcation type on the locations of the active nodes.
By choosing the coupling strength nearby the SW bifurcation $\varepsilon
=0.952\gtrsim \varepsilon _{2}$, we plot in Fig. 8(c) the node attaching
probability $p_{u2}$ as a function of the node index $i$. An interesting
finding is that, comparing to the situation of LW bifurcation [Fig. 8(a)],
the locations of the active nodes have been totally changed in Fig. (c). In
Tab. 2 we list the detail information about the $5$ most active nodes in
Fig. 8(c), again their locations can not be predicted by the node degree or
betweenness. Similar to the LW bifurcation, the locations of the active
nodes are also independent to the coupling strength at the SW bifurcation,
as shown in Fig. 8(d).

\begin{center}
\begin{table}[tbh]
\caption{Similar to Tab. I but for the attaching probability $p_{i}$ plotted
in Fig. 8(c). Comparing to Tab. I, one important observation is the changed
locations of the active nodes due to the changed bifurcation type.}%
\begin{tabular}{|c|c|c|c|c|c|c|}
\hline
Node index $i$ & $p_{i}$ & $p_{i}$\_rank & $k_{i}$ & $k_{i}$\_rank & $B_{i}$
& $B_{i}$\_rank \\ \hline
43 & 0.78196 & 1 & 9 & 779$\rightarrow $813 & 2847 & 813 \\ \hline
35 & 0.78969 & 2 & 18 & 936$\rightarrow $940 & 8513 & 953 \\ \hline
714 & 0.795 & 3 & 4 & 1$\rightarrow $338 & 1215 & 158 \\ \hline
130 & 0.79652 & 4 & 13 & 886$\rightarrow $901 & 4200 & 886 \\ \hline
154 & 0.19944 & 5 & 10 & 814$\rightarrow $846 & 2898 & 815 \\ \hline
\end{tabular}%
\label{table_SW}
\end{table}
\end{center}

Previous studies about network synchronization have shown that,
while individually it is difficult to predict the dynamical
behavior of each node, the average performance of an ensemble of
nodes of the same network properties do have some reliable
characters. For instance, it has been shown that in complex
networks the high-degree nodes are on average more synchronizable
than the low-degree ones \cite{ZK:2006}. Regarding to the problem
of node activities, it is natural to ask the similar question: are
the high-degree nodes more synchronized than the low-degree nodes?
In Fig. 9 we plot the average attaching probability $\left\langle
p_{u1}\right\rangle _{k}$ as a function of degree $k$. Still, we
can not find a clear dependence of $\left\langle
p_{u1}\right\rangle _{k}$ on $k$.
\begin{figure}[tbp]
\begin{center}
\epsfig{figure=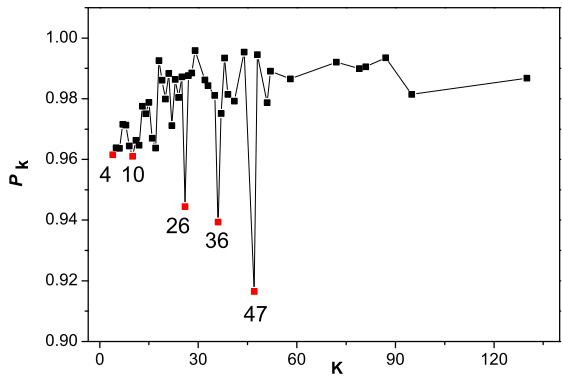,width=0.9\linewidth}
\end{center}
\caption{(Color online) The average attaching probability $\left\langle
p_{u1}\right\rangle _{k}$ as a function of node degree $k$. On average, the $%
5$ most unstable nodes are those of degrees $k=47,36,26,10,4$. Still, we can
not find a clear dependence between $\left\langle p_{u1}\right\rangle _{k}$
and $k$.}
\end{figure}

\section{Discussions and conclusion}

It is worthy of note that our studies of active nodes are only focused on
the giant-cluster state, and the purpose is to understand their dynamics and
reveal their properties. By ensemble average, we may able to improve our
prediction of the active nodes, say for example the dependence of $%
\left\langle p_{u1}\right\rangle _{k}$ on $k$ in Fig. 9 may be
smoothed if we average the results over a large number of network
realizations. Such an improvement, however, comes at the cost of
the decreased prediction accuracy due to the increased candidates.
Taking Fig. 9 as an example, although it is noticed that nodes of
$k=4$ in general are more active than those of other degrees, only
one of them is listed as the $5$ most unstable nodes [Tab. 1]. In
specific, among the total number of $338$ nodes which have degree
$k=4$, most of them are tightly attracted to the giant cluster
($90$ percent of them have attaching probabilities $p_{u1}>0.95$).
Therefore, in terms of precise predication, the average method is
infeasible in practice.

Beside node degree and betweenness, we have also checked the
dependence of the property of node activity to some other
well-known network properties such as the clustering coefficient,
the modularity, and the assortativity. However, none of them is
suitable to characterize the active nodes, their performance is
very similar to that of the node degree described in Tab. 1 and
Tab. 2. Our study thus suggests that, to give a precise prediction
to the active nodes, we may need to develop some new quantities.

Despite of the amount of studies carried on network synchronization, to the
best of our knowledge, we are the first to study the nonstationary pattern
in unsynchronizable complex networks. In Ref. \cite{SYN:DETECT1} the authors
have discussed the transient process of global synchronization in complex
networks, but their study are concentrating on the synchronizable state in
which, during the course of system evolution, small clusters integrate into
larger clusters monotonically and finally reach the synchronization state.
After that, the system will always stay on the synchronization state. Our
works are also different to the studies of Refs. \cite{SYN:DETECT2,SAH:2005}%
. Similar to our works, in these studies the authors also consider the
problem of pattern formation in unsynchronizable networks, but their
interests are focused on the \emph{stationary pattern} of the system. That
is, the size and contents of the clusters do not change with time. In
contrast, in our studies both the size and contents of the clusters are
variable.

In summary, we have reported and investigated a kind of new
phenomena in network synchronization: the nonstationary pattern.
That is, the final state of the network settles neither to the
synchronization state nor to any stationary state of fixed
pattern, the system is travelling among all the possible patterns
in an intermittent fashion (the pattern can be of any
configuration, but its probability of showing up is
pattern-dependent). We attribute this nonstationarity to the
asymmetric topology of the complex networks, and its dynamical
origin can be understood from the property of the finite-time
Lyapunov exponent associated to the desynchronized mode. Two types
of synchronization formats, the complete synchronization and the
temporal phase synchronization, have been employed to detect the
nonstationary dynamics. For coupling strength immediately out of
the stable region, the pattern evolution is characterized by the
process of on-off intermittency and the existence of the
giant-cluster; while if the coupling strength is far away from the
bifurcation points, the pattern evolution is signatured by the
random interactivities among the number of small clusters. A
remarkable finding is that, in the giant-cluster state the
locations of the active nodes are independent of the coupling
strength but are sensitive to the bifurcation types. The active
nodes, however, can not be characterized by the currently known
network properties, further investigations about their
identification are necessary. While we are hoping our studies
about nonstationary pattern could give some new understandings to
the dynamics of coupled complex systems, we also hope that our
findings about unsynchronizable networks could be used to some
practical problems where system maintains their normal functions
only under the unsynchronizable states, for example the problem of
epileptic seizers \cite{LFOH:2007}.

\end{document}